\documentclass[11pt]{article}
\usepackage{graphicx,amsmath,amssymb}
\usepackage[dvipsnames]{xcolor}
\usepackage[a4paper,top=2.2cm,bottom=2.2cm,left=2.2cm,right=2.2cm]{geometry}
\usepackage[colorlinks=true,linkcolor=blue,citecolor=blue,urlcolor=blue]{hyperref}

\usepackage{lineno}

\newcommand{\KS}{\ensuremath{K^0_S}}
\newcommand{\KL}{\ensuremath{K^0_L}}
\newcommand{\Kz}{\ensuremath{K^0}}
\newcommand{\Kzb}{\ensuremath{\bar K^0}}
\newcommand{\dphio}{\ensuremath{\Delta\phi_{\rm o}}}

\title{Odderon exchange in high-energy $K^0_S$ regeneration at the LHC}

\author{P.~Filip$^{1,2}$, M.~Ta\v sevsk\'y$^{1}$, V.~A.~Khoze$^{3}$, R.~Pasechnik$^{4}$, M.~G.~Ryskin$^{5}$ and B.~G.~Zakharov$^{6}$\\[3mm]
{\normalsize\itshape $^{1}$Institute of Physics of the Czech Academy of Sciences, Prague, Czech Republic}\\
{\normalsize\itshape $^{2}$Institute of Physics, Slovak Academy of Sciences, Bratislava, Slovakia}\\
{\normalsize\itshape $^{3}$Institute for Particle Physics Phenomenology, University of Durham, UK}\\
{\normalsize\itshape $^{4}$Department of Physics, Lund University, Lund, Sweden}\\
{\normalsize\itshape $^{5}$Petersburg Nuclear Physics Institute, NRC ``Kurchatov Institute'', Gatchina, St.~Petersburg, Russia}\\
{\normalsize\itshape $^{6}$L.D.~Landau Institute for Theoretical Physics, Chernogolovka, Russia}}

\date{}

\begin{document}
\maketitle

\begin{abstract}
\noindent
We revisit the possibility of detecting the Odderon exchange
through high-energy neutral-kaon regeneration, focusing on the
in-matter $\KL\to\KS$ conversion of $(\sim\!0.2$--$2)$~TeV \KL{}
mesons originating from $pp$ collisions at $\sqrt{s}=13.6$~TeV, and on the
practical constraints of realizing such a measurement in the very-forward
region of an LHC interaction point.
The analysis has two complementary parts.  First, we reproduce the original
coherent-forward-regeneration estimates for a liquid-hydrogen regenerator
and extend them to realistic C, Cu and Pb regenerators of an LHC-compatible
geometry, including neutral-kaon attenuation.  We show that an
Odderon-induced regeneration phase produces a measurable distortion of the
$\Kz\to\pi^0\pi^0$ decay-vertex distribution at TeV kaon energies, but that
the survival of primary \KS{} mesons from the interaction point imposes
severe baseline requirements, while a competing electromagnetic $C=-1$
(photon-exchange) amplitude limits the interpretation of the coherent mode
as a clean Odderon measurement.  Second, we examine non-forward (diffractive)
regeneration at lower kaon energies of $0.2$--$0.8$~TeV, where the
primary-\KS{} contamination is naturally suppressed, and estimate the
competing Odderon, Pomeron--Odderon-cut, $\omega$-Reggeon and
photon-exchange contributions to the regeneration amplitude.  We identify neutron-induced neutral-only
strangeness production and inelastic Regge backgrounds as the dominant
limitations, quantify the background suppression required for an observable
Odderon signal, and formulate the ingredients of an active double-regenerator
subtraction strategy.
\end{abstract}

\section{Introduction}
\label{sec:intro}
Apart from the even-signature singularity (Pomeron), in QCD with $N_c=3$
there exists its counterpart, the odd-signature singularity placed at
$j\simeq 1$.  It is formed by three $t$-channel reggeized gluons connected in
colour space by the symmetric $d^{abc}$ tensor of the colour $SU(3)$ group,
see Section~20 in Ref.~\cite{PDG}. This contribution to the amplitude is called the Odderon.
The existence of a crossing-odd high-energy singularity was first introduced
in the context of asymptotic theorems by Lukaszuk and
Nicolescu~\cite{LukNic}, and the name ``Odderon'' was subsequently attached
to this contribution in Ref.~\cite{JLLN}.  Its observation remains
non-trivial because the dominant high-energy elastic amplitude is $C$-even
and predominantly imaginary.

The Odderon exchange amplitude has the opposite sign in proton--proton and
proton--antiproton scattering.  Since the Odderon intercept is very close to
$\alpha_{\rm Odd}(0)=1$, the corresponding amplitude grows approximately
proportionally to $s$; hence its contribution {\it relative} to the dominant
$C$-even (Pomeron) amplitude, and thus to the cross section, does not
decrease, or decreases only very slowly, with energy.
According to perturbative estimates, the coupling of the Odderon to the
nucleon is rather small.  The corresponding amplitude is mainly real and is
about 100~times smaller than the imaginary part of the Pomeron exchange
amplitude, see the recent review~\cite{Ryskin} and references therein.
Therefore, extracting the Odderon contribution from the data on top of a
much larger $C$-even contribution is very challenging.

The conventional Odderon searches are based on studies of $pp$ and $p\bar p$
scattering in the very low $|t|$ region of Coulomb--nuclear interference, or
in the diffractive dip region where the imaginary part of the $C$-even
amplitude vanishes.

Recall that, due to dispersion relations, the real part of the high-energy
$C$-even amplitude is relatively small
(Re\,$A_{\rm even}\ll$\,Im\,$A_{\rm even}$).  The major constraint on the
$C$-odd amplitude is that both the particle and the antiparticle cross
sections must be positive, while the $C$-odd amplitude changes its sign from
particle to antiparticle.  This condition must be satisfied at any energy
and at each impact parameter, $b$, that is, at any partial wave
$l=b\sqrt{s}/2$.

This means that for the intercept and the $t$-slope of the Odderon
trajectory
\begin{equation}
\label{eq:odd-bounds}
\alpha_{\rm Odd}(0)<\alpha_{\rm even}(0)
\quad\mbox{and}\quad
B_{\rm Odd}<B_{\rm even}\,.
\end{equation}
Perturbative QCD satisfies these conditions: for the three-gluon diagram one
gets $\alpha_{\rm Odd}(0)=1$, while the Pomeron intercept is
$\alpha_{\rm Pom}(0)>1$.  The real part of the proton--proton amplitude can
be measured via the Coulomb--nuclear interference at very low momentum
transfer $|t|$.  In 2018--2020, the TOTEM collaboration at the LHC claimed
the Odderon discovery based on two results.  First, TOTEM measured elastic
$pp$ scattering at low $|t|$, down to $-t=8\cdot 10^{-4}$~GeV$^2$, and
determined the real-to-imaginary-part ratio
$\rho=\mbox{Re}/\mbox{Im}\simeq 0.09$--$0.10$, which turned out to
be~\cite{rho-T} noticeably smaller than the expected value
($\rho=0.13$--$0.14$) coming from dispersion relations for the case of pure
$C$-even interactions.

TOTEM has also measured the $pp$ cross section in the diffractive dip region
at $\sqrt{s}=2.76$~TeV~\cite{2.76}.  A clear `dip--bump' structure was
observed, while at the relatively close Tevatron energy $\sqrt{s}=1.96$~TeV
the $t$~dependence of the $\bar pp$ cross section is more or less
flat~\cite{D0-T}.  This was interpreted as the presence of the Odderon real
part, which diminishes the real part of the $pp$ amplitude but enlarges it,
filling the dip, in the $\bar pp$ case\footnote{Ideally, it would be crucial
to have analogous measurements of $d\sigma_{\rm el}/dt$ around the dip in
both proton--proton and proton--antiproton elastic scattering at the same
energies (better still, with the same apparatus).}.

The real part of the $C$-even amplitude can be calculated from dispersion
relations.  At high energies it takes the form
\begin{equation}
\label{eq:disp}
\rho_{\rm even}\simeq \frac{\pi}{2}\,
\frac{\partial\ln \sigma_{\rm tot}(s)}{\partial\ln s}\ .
\end{equation}
That is, the value of $\rho$ is strongly correlated with the energy
behaviour of the total cross section.  Using the cross sections measured by
TOTEM without the Odderon, one expects $\rho=\rho_{\rm even}=0.13$--$0.14$
and not the $\rho=0.09$--$0.10$ observed by TOTEM~\cite{rho-T}.  However,
the normalisation used by TOTEM could be questioned.  Note that the values
of $\sigma_{\rm tot}$ measured by ATLAS-ALFA at 7, 8 and 13~TeV are found to
be systematically lower than the TOTEM numbers.  In particular, the 13~TeV
ATLAS data give the same $\rho$ as TOTEM, $\rho=0.10\pm 0.01$, but the value
of the total cross section is
$\sigma_{\rm tot}=104.68\pm 1.09$~mb~\cite{atl13}, which is approximately
5\% lower than the average of the values determined by TOTEM
($\sigma_{\rm tot}=110.6\pm 3.4$~mb, $109.5\pm 3.4$~mb and
$110.3\pm 3.5$~mb), indicating that a smaller value of the real part of the $C$-even 
amplitude should be expected from the dispersion relations.  The relatively small
value of $\rho$ can then be explained by an admixture of the $C$-odd
amplitude, which survives at high LHC energies.

The available low-$|t|$ ($|t|<0.1$~GeV$^2$) data at
$50~\mbox{GeV}<\sqrt{s}<13$~TeV were analysed in the recent
paper~\cite{LRK}, including both the TOTEM and ATLAS-ALFA results with the
corresponding (free) normalisation factors.  A quite satisfactory fit was
obtained, with $\chi^2=560$ for $\nu=504$ degrees of freedom,
i.e.\ $\chi^2/\nu=1.11$.  Neglecting the Odderon would lead to a much larger
$\chi^2=726$.  The main conclusions of the study in Ref.~\cite{LRK} are:
\begin{itemize}
\item the description with the addition of the Odderon improves the fit
      (with the Odderon, $\chi^2$ becomes much smaller);
\item the sign of the Odderon amplitude needed to describe the very low
      $|t|$ data is opposite to that predicted by the perturbative QCD
      three-gluon exchange contribution\footnote{The problem can be solved
      assuming that the Odderon coupling $\beta_{\rm o}$ vanishes (or
      decreases strongly) at $t=0$.  In this case, the dominant $C$-odd
      contribution at $t=0$ comes from the Pomeron--Odderon cut and has the
      opposite sign~\cite{Zakh}.};
\item the Odderon--proton coupling, $\beta_{\rm o}$, is smaller than that of
      the Pomeron.  Moreover, after accounting for the screening of the seed
      Odderon by the Pomeron cuts, the final $C$-odd contribution to $\rho$
      at 13~TeV becomes quite small,
      $\delta\rho=(\rho^{\bar pp}-\rho^{pp})/2\leq 0.004$, i.e.\ ten~times
      smaller than that ($\delta\rho=0.04$) originally claimed by TOTEM.
\end{itemize}

The TOTEM publications~\cite{rho-T,2.76} prompted a renewal of interest in
establishing the high-energy $C$-odd (Odderon) contribution.  Though the
existence of the Odderon is a firm prediction of QCD, further efforts are
certainly needed to confirm its presence experimentally and to measure its
coupling to the proton accurately.  In the literature, different approaches
have been proposed to reveal the presence of the Odderon, see the
reviews~\cite{PDG,Ryskin}.

This motivates searches for independent observables in which the Odderon enters a
different amplitude combination from that probed by elastic $pp$ and
$p\bar p$ scattering.  Neutral-kaon regeneration provides such a
possibility, because the regeneration amplitude is sensitive to the
{\it difference} between the \Kz{} and \Kzb{} forward scattering amplitudes
in matter, and hence directly to crossing-odd exchange contributions
(including, as discussed below, the electromagnetic photon-exchange
amplitude, which must be controlled).  In the eighties it was suggested that
the Odderon exchange could be detectable using \KS{} regeneration, see the
original papers in Refs.~\cite{OddKiev,OddZach}.  Here we revisit this
independent method in more detail: we consider high-energy neutral kaons
produced in $\sqrt{s}=13.6$~TeV $pp$ collisions at the LHC and discuss the
possibilities of extracting the Odderon signal via the regeneration of
\KS{} mesons resulting from the interaction of \KL{} mesons with nucleons
and nuclei.

\section{Introduction to the regeneration process}
\label{sec:regen}
By {\it coherent forward regeneration}\footnote{Throughout this paper,
``forward'' in ``coherent forward regeneration'' refers to the (near-)zero
scattering angle of the $\KL\to\KS$ conversion in the regenerator; it should
not be confused with the very-forward production region of the $pp$
collision (pseudorapidities $\eta\gtrsim 8$), in which all neutral kaons
considered here are emitted.} we understand such \KL{} interactions
with the regenerator nuclei (or electrons) which result in a \KS{} neutral
meson state with the same momentum $\vec p$ as the original incoming \KL{}.
This means $|t|\rightarrow t_{\rm min}$ and scattering angle
$\theta\approx 0$.  In this case the interference $|A_1+A_2|^2$ of the
amplitudes $A_1(\KL\rightarrow\pi\pi)$ and
$A_2(\KL\rightarrow\KS\rightarrow\pi\pi)$ occurs, because the momenta of the
original \KL{} and of the regenerated \KS{} coincide, resulting in an
identical final state $\pi\pi$.  Scattering centres (nuclei) in the
regenerator add coherently to the total amplitude of such a
process~\cite{CohRegenerMain,CohRegenNEW}.

By {\it non-forward regeneration} of \KS{} from the incoming \KL{} mesons we
understand the process in which the momentum $\vec p_s$ of the outgoing
\KS{} meson acquires a transverse component that measurably differs from the
momentum $\vec p_L$ of the incoming \KL{} particle, with the scattering
angle $\vec p_L\cdot\vec p_s/(|\vec p_L|\,|\vec p_s|)=\cos\theta<1$.  This
means that the final $\pi\pi$ state produced in the direct CP-violating
decay $\KL\rightarrow\pi\pi$ differs from the final state of the indirect
chain $\KL\rightarrow\KS\rightarrow\pi\pi$, and thus the amplitudes of these
processes do not interfere.  The contributions of the scattering centres
(nuclei) in the regenerator do not add coherently in this case (unless the
momentum exchange of the regenerated \KS{} meson with the nuclei is so small
that a non-negligible fraction of the nuclei, determined by the Debye--Waller
factor~\cite{D-W_factor,D-W_Lipkin}, still interacts coherently).

\subsection{Regeneration amplitude, attenuation and the Odderon-sensitive
phase}
\label{sec:regen-amp}
For later use, let us make explicit the normalisation of the
coherent-regeneration amplitude.  In the thin-regenerator limit the forward
conversion amplitude is controlled by the difference of the forward
scattering amplitudes of the neutral-kaon flavour eigenstates in matter.  Up to an
overall convention-dependent phase,
\begin{equation}
\label{eq:rhoA}
\varrho_A \simeq -\,\frac{i\pi n_A L_A}{p_K}\,
\big[f_{\Kzb A}(0)-f_{\Kz A}(0)\big]\,,
\end{equation}
where $f_{\Kz A}(0)$ and $f_{\Kzb A}(0)$ denote the forward elastic
scattering amplitudes of the \Kz{} and \Kzb{} flavour eigenstates on the
target species $A$, $n_A$ is the number density of scattering centres, $L_A$ is the regenerator
length (its extension along the beam) and $p_K$ is the kaon momentum; only phase differences between the
regenerated amplitude and the direct $\KL\to\pi\pi$ amplitude are
observable.  Equation~(\ref{eq:rhoA}) is the thin-target form used only to
define conventions; for a thick regenerator the finite propagation and the
$\KS$--$\KL$ eigenvalue splitting must be included through the
optical-potential evolution.  With this convention, the measurable
regeneration phase is $\phi=\arg\varrho_A$, while an Odderon contribution
modifies both $|\varrho_A|$ and $\phi$.  More generally, finite thickness
and absorption can be included by propagating the two-component
$(\Kz,\Kzb)$ or $(\KS,\KL)$ system through the optical potential
\begin{equation}
\label{eq:optical}
U_{K,\bar K} = -\,\frac{2\pi n_A}{E_K}\,f_{K,\bar K;A}(0)\,,
\qquad
\psi(L_A)=\exp\big[-iH_{\rm eff}L_A\big]\,\psi(0)\,,
\end{equation}
or, equivalently for the present estimates, by multiplying the generated
\KS{} and surviving \KL{} components by the appropriate attenuation factors
\begin{equation}
\label{eq:attenuation}
A_i(L_A)=\exp\big[-n_A\,\sigma^{\rm abs}_i(E_K,A)\,L_A\big]\,,
\qquad i=\KL,\KS\,.
\end{equation}
In the numerical curves shown below in Fig.~\ref{fig:Our_prediction} we use a common neutral-kaon attenuation
factor whenever the available input is not precise enough to warrant a separate
$\KL/\KS$ treatment. The Odderon phase shifts quoted below should therefore
be understood as benchmark distortions of the regeneration phase rather than
as a precision extraction from present elastic-scattering fits.  A future
experimental proposal would need to replace these benchmark phases by a
global model in which $f_{\Kzb A}(0)-f_{\Kz A}(0)$ is constrained
simultaneously by the existing regeneration data, high-energy hadron
scattering and nuclear attenuation.

\section{Coherent forward \boldmath$K^0_S$ regeneration at TeV energies}
\label{sec:coherent}
High-energy \KS{} regeneration ($E_{K}\leq 130$~GeV) was studied at
Fermilab, where indications of an unexpected energy dependence of the
regeneration parameters were reported~\cite{Aronson-FNAL}.  In
Refs.~\cite{OddKiev,WinsteinPhaseShift} the authors conclude that the
Odderon exchange manifests itself through a change of the \KS{} coherent
regeneration phase ($\phi$) with energy, which consequently modifies the
intensity of the \Kz{} decays behind the regenerator.  As shown in Fig.~5 of
Ref.~\cite{OddKiev}, the Odderon influence is expected to be negligible for
energies $E_{K}\lesssim 0.5$~TeV but significant at 1~TeV and higher
energies.

\begin{figure}[t!]
\begin{center}
 \includegraphics[width=0.72\textwidth]{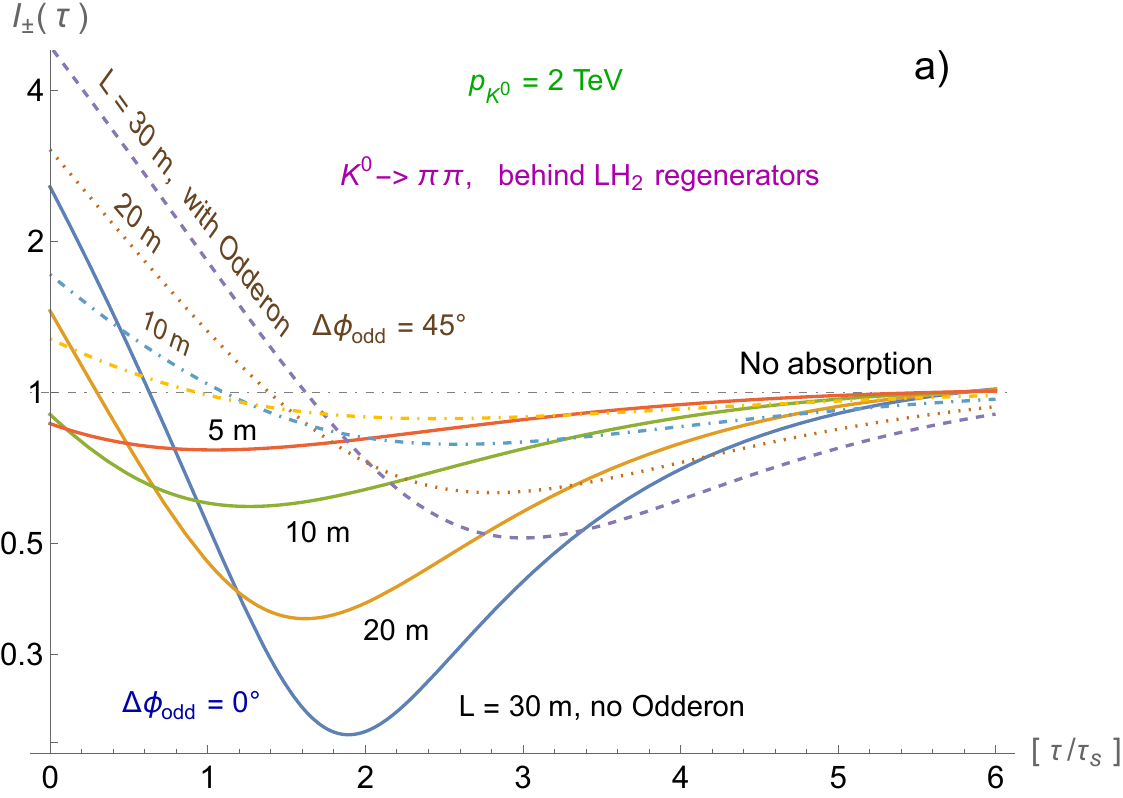}
\caption{The ratio of the $\pi\pi$ decay intensity behind the regenerator to
that without a regenerator, $I_{\pm}(\tau)$, for an LH$_2$ regenerator of
several lengths.  The ratio is shown for 2~TeV kaons as a function of the
kaon proper time $\tau$ in units of the \KS{} mean lifetime $\tau_S$,
without the \Kz{} absorption in the LH$_2$ regenerator (our curves are to be
compared with Fig.~6 of Ref.~\cite{OddKiev}).}
\label{fig:Kiev_reproduction}
\end{center}
\end{figure}

\begin{figure}[t!]
\begin{center}
 \includegraphics[width=0.72\textwidth]{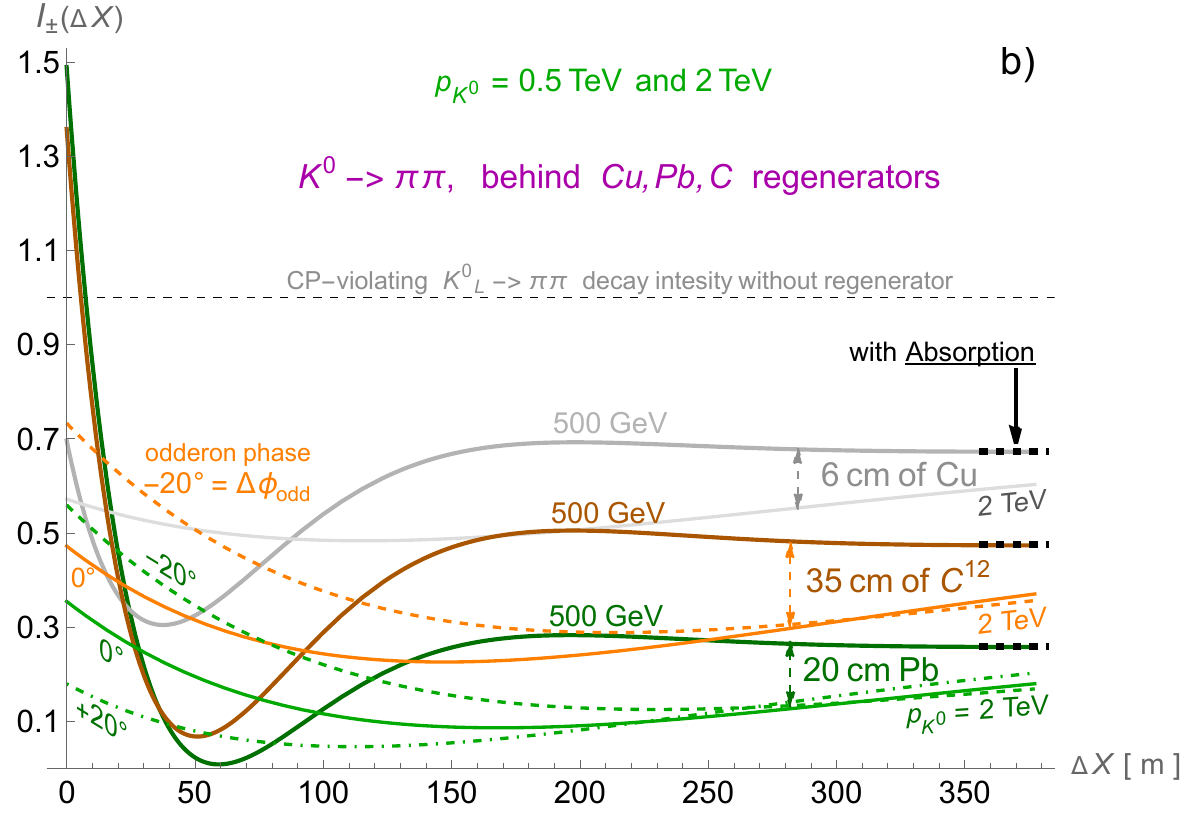}
\caption{The ratio of the $\pi\pi$ decay intensity with a regenerator to
that without a regenerator, $I_{\pm}(\Delta X)$, calculated for 500~GeV and
2~TeV kaons as a function of the distance of the $\pi\pi$ decay vertex from
the regenerator, for Cu, Pb and C regenerators, with the effect of \Kz{}
absorption included.  The ratios without (with) the Odderon exchange are
shown by the full (dashed) curves.}
\label{fig:Our_prediction}
\end{center}
\end{figure}

As a first step, we have reproduced the published~\cite{OddKiev}
intensities $I_{\pm}(\tau)$ of $\Kz\rightarrow\pi\pi$ decays, where $\tau$
is the proper decay time of the \Kz{} after leaving an LH$_2$ regenerator of
a length varied between 5 and 30~m (see Fig.~\ref{fig:Kiev_reproduction},
to be compared with Fig.~6 of Ref.~\cite{OddKiev}), by fitting the curves of
Ref.~\cite{OddKiev} with the interference formula
\begin{equation}
\label{eq:mainF}
I_{\pm}(\tau) = I_0\, B^K_{\pi\pi}
\Big[\,|\varrho|^2 e^{-\tau/\tau_S} + |\eta|^2 e^{-\tau/\tau_L}
 + 2|\varrho||\eta|\,
   e^{-\frac{\tau}{2}\left(\frac{1}{\tau_S}+\frac{1}{\tau_L}\right)}
   \cos\!\big(\Delta\omega_{LS}\,\tau + \phi + \dphio\big)\Big]\,.
\end{equation}
Introducing the $\KL$--$\KS$ mass difference, 
$\Delta m_K \equiv m_{\KL}-m_{\KS}$, the corresponding oscillation 
angular frequency $\Delta\omega_{LS}$ reads
\begin{equation}
\label{eq:omegaLS}
\Delta\omega_{LS} \equiv \frac{\Delta m_K c^2}{\hbar}
 = 5.293\cdot 10^{9}\,{\rm s}^{-1}\,.
\end{equation}
Equivalently, in the PDG convention one quotes
$\Delta m_K c^2 = 5.293\cdot 10^{9}\,\hbar\,{\rm s}^{-1}$.  With this
definition the argument of the cosine in Eq.~(\ref{eq:mainF}) is
dimensionless, and the numerical input can be read directly from the
standard neutral-kaon data tables. The values of the other (not fitted) parameters are
$\tau_S=8.954\cdot 10^{-11}$~s (the \KS{} mean lifetime),
$|\eta|=2.22\cdot 10^{-3}$, and $\tau_L/\tau_S=571.4$, the ratio of the
\KL{} to \KS{} mean lifetimes.  For the purpose of this reproduction the
absorption of neutral kaons in the LH$_2$ regenerator was not considered,
following Ref.~\cite{OddKiev}.

The fit of the curves in Fig.~6 of Ref.~\cite{OddKiev} returns an effective
Odderon phase shift $\dphio=45^{\circ}$ together with a small
($\Delta|\varrho|/|\varrho|\simeq 5\%$) increase of the regeneration
amplitude with respect to the no-Odderon parametrisation.  These values
coincide with the Odderon-induced modifications which the authors of
Ref.~\cite{OddKiev} obtained from their own model at $E_K\simeq 2$~TeV
(their Figs.~4 and~5) and then built into their illustrative decay curves.
Our fit therefore validates the normalisation and phase conventions of
Eq.~(\ref{eq:mainF}); it must not be interpreted as an independent
determination of the Odderon phase, nor as a claim that a $45^{\circ}$
phase is preferred by present data.
Notably, the same calculation predicts a considerably more moderate effect,
$\dphio\approx 20^{\circ}$, at UNK energies $E_K\approx 1$~TeV, in line
with the earlier estimates of the same mechanism~\cite{OddZach,ZaP1}.
For the phenomenological illustrations below we therefore adopt
$|\dphio|=20^{\circ}$ as a conservative benchmark for the Odderon-induced
distortion, while the $45^{\circ}$ curves in
Fig.~\ref{fig:Kiev_reproduction} serve only to reproduce
Ref.~\cite{OddKiev}.  We stress that the Odderon-induced phase shift is not
a universal constant: for a $C$-odd Regge singularity the phase is governed
by the signature factor, and hence by the Odderon intercept, so that
$\dphio$ in general depends on both the collision energy and the momentum
transfer.  The constant values quoted above are to be understood as
effective benchmarks at fixed $E_K$ in the forward limit.

In Fig.~\ref{fig:Our_prediction}, the full curves show our prediction for
the $\Kz\rightarrow\pi\pi$ decay intensity without the Odderon exchange for
0.5~TeV and 2~TeV kaons (this time with absorption included) behind 6~cm of
Cu, 35~cm of C, or 20~cm of Pb, as a function of the distance of the decay
vertex from the regenerator.  For the chosen lengths of the C and Pb
regenerators, the \KS{} regeneration dominates over the \KL{} absorption,
and consequently the total intensity of 0.5~TeV $\Kz\rightarrow\pi\pi$
decays exceeds the intensity of the CP-violating $\KL\rightarrow\pi\pi$
decays without a regenerator (within 10--20~m behind the regenerator).  The
total intensity then shows a dip, behind which the number of \KS{} decays
rises up to the point where all regenerated \KS{} mesons have decayed and
the absorption-determined plateau is reached.

The effect of the Odderon exchange is demonstrated by the dashed curves in
Fig.~\ref{fig:Our_prediction}.  As explained above, its effect is negligible
for kaon energies of 0.5~TeV and below (the dashed counterparts of the
0.5~TeV full curves are not shown), but a significant effect is observed for
the 2~TeV case: the Odderon phase shift $|\dphio|=20^{\circ}$ modifies the
$\Kz\rightarrow\pi\pi$ intensity curves in an observable way.  For such a
detection of the Odderon exchange, the $\Kz\rightarrow\pi\pi$ decay-vertex
position and the \Kz{} energy need to be reliably measured.  We believe that
an upgraded LHCf-type detector~\cite{LHCf} is capable of such measurements and
conclude that its suitable location would be from about 20 to 150~m behind
the regenerator (where the differences between the full and dashed curves
are significant).  The selected lengths, $L_{\rm Cu}=6$~cm,
$L_{\rm C}=35$~cm and $L_{\rm Pb}=20$~cm, should be regarded as benchmark
choices balancing the regeneration probability against absorption.  In a
final detector design these choices should be re-optimised using a
material-by-material figure of merit, for example
\begin{equation}
\label{eq:FoM}
F_A(E_K,L_A)=\frac{|N_A(\dphio)-N_A(0)|}{\sqrt{N_A(0)+B_A}}\,,
\end{equation}
where $N_A(\dphio)-N_A(0)$ denotes the expected change of the reconstructed
$\Kz\rightarrow\pi^0\pi^0$ yield in the chosen decay-volume window when the
Odderon phase is switched on, while $B_A$ denotes the sum of the
primary-\KS{}, neutron-induced and detector backgrounds.  This quantity is
not used as an additional numerical result here; it only indicates the
signal-to-statistical-uncertainty criterion to be maximised in a future
detector optimisation.

The regeneration amplitude is sensitive to {\it any} $C=-1$ exchange, and
photon exchange therefore contributes as well, through the interaction of
the neutral kaon with both the electrons and the protons of the target.  In
the $f(0)$ approximation for a hydrogen target, the direct electron and
proton photon-exchange terms cancel each other, and the residual
electromagnetic contribution is dominated by the absorptive
Pomeron$\otimes\gamma$ term.  For nuclear targets, Glauber screening reduces
the effective number of active protons,
$Z_{\rm eff}/Z\approx 0.25\,(0.6)$ for heavy (light) nuclei, and the
electromagnetic part of the forward amplitude becomes
\begin{equation}
\label{eq:femA}
f^{\rm em}_A(0)\;\approx\;
Z_{\rm eff}\, f^{\,p}_{P\otimes\gamma}(0)\;-\;
\big(Z-Z_{\rm eff}\big)\, f^{\,p}_{\gamma}(0)\,.
\end{equation}
For the effective gluon mass $m_g=0.17$~GeV (see
Section~\ref{sec:nonforward}) one finds
$|f^{\rm em}_A(0)/f^{\,P\otimes {\rm Odd}}_A(0)|\sim 0.5$--$2$, depending on
the form-factor parametrisation and on the quasi-eikonal parameter $C$,
while for $m_g=0.7$~GeV the electromagnetic term dominates.  The \Kz{}
electromagnetic form factor entering these estimates is evaluated in a
constituent-quark picture~\cite{GNS}, normalised to the measured $K^+$
charge radius $\langle r^2_{K^+}\rangle\approx 0.34$~fm$^2$~\cite{Amendolia};
the atomic form factors are treated in the Gribov formalism~\cite{Gribov}.
Consequently, coherent regeneration alone cannot be regarded as a clean
Odderon extraction unless the electromagnetic $C=-1$ amplitude is computed
and constrained to sufficient accuracy.  In principle, targets with
different $Z/A$ ratios (say, hydrogen and deuterium) can be used to
disentangle the photon and Odderon contributions.  We also note that the
electromagnetic contribution may account for at least part of the
long-standing difference between the values of $\alpha_\omega(0)$ extracted
from the energy dependence of the coherent regeneration amplitude on
protons~\cite{Bock} and on nuclei~\cite{Gsponer}.  A detailed account of
these electromagnetic estimates will be presented elsewhere.

\section{Non-forward \boldmath$K^0_S$ regeneration at 200--800 GeV energies}
\label{sec:nonforward}
\begin{figure}[t!]
\begin{center}
\includegraphics[width=0.8\textwidth]{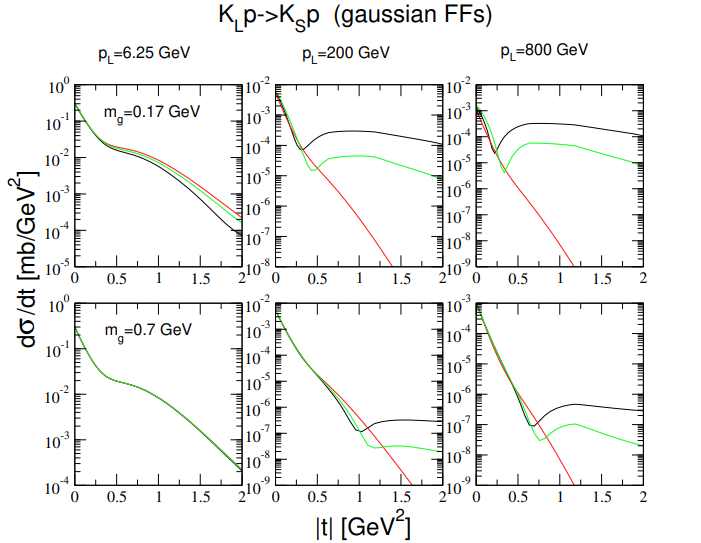}
\caption{Predictions for the non-forward \KS{} regeneration cross section on
protons for three values of the \KL{} momentum and two values of the
effective gluon mass, using Gaussian form factors.  The red curve shows the
cross section due to the $\omega$-Reggeon contribution alone; the black and
green curves show the predictions for the full amplitude
$(T_{\rm Odd}+T_{P\otimes{\rm Odd}}+T_R)$ for $C=1$ and $C=1.8$,
respectively.}
\label{fig:nonfwdG}
\end{center}
\end{figure}

In terms of the three-gluon exchange, the Odderon amplitude can be estimated
in the Born approximation with running $\alpha_s$,
\begin{equation}
\label{eq:TOdd}
T_{\rm Odd}(Q)=\frac{10\,s}{27\pi}\int d^2q_1\, d^2q_2
\prod_{i=1}^{3}\frac{\alpha_s(q_i^2)}{(q_i^2+m_g^2)}\,
R_N(q_1,q_2,q_3)\, R_K(q_1,q_2,q_3)\,,
\end{equation}
with $q_3=Q-q_1-q_2$ and $R_N$, $R_K$ defined in Ref.~\cite{ZaP1}, using
$R_{N,K}=V_{N,K}$ with $V_{N,K}$ from Ref.~\cite{OddKiev}.  We use the
one-loop running $\alpha_s(Q^2)$ frozen at the value $\alpha_s^{fr}$ for
$Q<Q_{fr}=\Lambda_{\rm QCD}\exp\big(2\pi/9\alpha_s^{fr}\big)$.  The
parameter $\alpha_s^{fr}$ is adjusted to satisfy the
Dokshitzer--Khoze--Troyan (DKT) infrared condition~\cite{ZaP3}
\begin{equation}
\label{eq:DKT}
\int_0^{2\,{\rm GeV}}\frac{dq}{\pi}\,\alpha_s(q^2)\;\approx\;0.36\ {\rm GeV}\,,
\end{equation}
which gives $\alpha_s^{fr}\approx 1.04$ (for
$\Lambda_{\rm QCD}=200$~MeV).
The scale argument in Eq.~(\ref{eq:TOdd}) is chosen as $q_i^2$ for each
exchanged gluon; varying this prescription is part of the
perturbative-model uncertainty, together with the effective gluon mass and
the hadronic form factors.  Similarly to Ref.~\cite{ZaP1}, we calculate the
contribution of the Pomeron--Odderon cut in the quasi-eikonal approximation
(for $C=1$ and $C=1.8$):
\begin{equation}
\label{eq:PxO}
T_{P\otimes{\rm Odd}}(Q)=\frac{iC}{8\pi^2 s}\int d^2q\;
T_P(q)\,T_{\rm Odd}(Q-q)\,.
\end{equation}
The parameter $C$ accounts phenomenologically for the quasi-eikonal
enhancement of intermediate diffractive states; we show both $C=1$ and
$C=1.8$ in order to display the uncertainty associated with the absorptive
corrections. (The symbol $C$ is used here for this quasi-eikonal parameter
only, following Ref.~\cite{ZaP1} and the labels of
Figs.~\ref{fig:nonfwdG}, \ref{fig:nonfwdP} and \ref{fig:Zakh3}; it should
not be confused with the charge parity $C=\pm 1$.)  The two values of the effective gluon mass, $m_g=0.17$ and
$0.7$~GeV, should similarly be read as a deliberately broad range: the
latter is close to the inverse non-perturbative gluon-correlation length in
the QCD vacuum, while the former reproduces the more infrared-sensitive
benchmark of the older calculation~\cite{OddKiev}, with $m_g$ of the order
of the inverse proton radius (note that it is of the order of the infrared
regulator $m\sim 0.1$--$0.3$~GeV which is often used in calculations within
the Colour Glass Condensate picture~\cite{CGC1,CGC2}).

Similarly to Ref.~\cite{ZaP1}, the $\omega$-Reggeon amplitude $T_R$ is
calculated within the Harari model~\cite{Hara4,Hara5}, using
$T_R^{nf}(s,t)$ in the form given by Ref.~\cite{ZaP6},
\begin{equation}
\label{eq:TRnf}
T_R^{nf}(s,t)=g_0\,(s/s_0)^{\alpha(t)}\, e^{tA}
\big[(1+at+bt^2)\,e^{tB}\tan\!\big(\pi\alpha(0)/2\big)+iJ_0(r\sqrt{-t})\big]
\end{equation}
and
\begin{equation}
\label{eq:TRsf}
T_R^{sf}(s,t)=g_1\,(s/s_0)^{\alpha(t)}\, e^{tA}\,J_1(r\sqrt{-t})
\big[\tan\!\big(\pi\alpha(0)/2\big)+i\big]\,,
\end{equation}
with $r=5.19$~GeV$^{-1}$, $A=-0.88$~GeV$^{-2}$, $g_0=-19$, $g_1=17$,
$a=2.79$~GeV$^{-2}$, $b=8.78$~GeV$^{-4}$, $B=82$~GeV$^{-2}$ and
$\alpha(t)=\alpha(0)+t\alpha'$, where $\alpha(0)=0.43$ and
$\alpha'=0.88$~GeV$^{-2}$.

Spin-flip Odderon and Pomeron vertices can arise from the
transverse-momentum components of the light-cone nucleon wave function:
although the gluon--quark interaction conserves helicity at high energies,
the sum of the quark helicities need not coincide with the nucleon helicity.
Such vertices are expected to give only a small correction to the
unpolarised $d\sigma/dt$ and are neglected here, together with the
corresponding photon spin-flip terms.
The $\KL+p\rightarrow\KS+p$ amplitude then reads
\begin{equation}
\label{eq:Ttot}
T=T_{\rm Odd}+T_{P\otimes{\rm Odd}}+T_R\,.
\end{equation}
Since the Pomeron amplitude is dominantly imaginary at high energies,
Eq.~(\ref{eq:PxO}) gives a dominantly {\it real} Pomeron--Odderon cut; the
small imaginary correction induced by ${\rm Re}\,T_P$ is neglected in the
numerical estimates.  The differential cross section for \KS{} regeneration
then reads
\begin{equation}
\label{eq:dsigdt}
\frac{d\sigma}{dt}=N_T\Big[
\big(T_{\rm Odd}+{\rm Re}\,T_{P\otimes{\rm Odd}}+{\rm Re}\,T_R^{nf}\big)^2
+\big({\rm Im}\,T_R^{nf}\big)^2+\big|T_R^{sf}\big|^2\Big]\,,
\end{equation}
where $N_T$ denotes the amplitude-normalisation factor, including the
conversion to mb/GeV$^2$ used in the plots; with the standard invariant
high-energy convention, $N_T=1/(16\pi s^2)$ before unit conversion, and the
numerical curves in Fig.~\ref{fig:nonfwdG} and in the Appendix use the same
normalisation convention as Ref.~\cite{ZaP1}.  If a different amplitude
convention is adopted, only the absolute vertical normalisation, not the
relative shape or the interference pattern, is affected.
We perform the calculation for the {\it Gaussian}
($F_i(Q)=\exp(-R_i^2Q^2)$, $i=p,K$) and {\it pole}
($F_i(Q)=1/(1+R_i^2Q^2)$, $i=p,K$) parametrisations of the proton and \Kz{}
form factors.  The spread between these two choices should be interpreted as
a model uncertainty of the non-forward prediction, not as a statistical
error.  This uncertainty is particularly important near the diffractive
minimum, where the relative size and sign of the $\omega$- and
Odderon-induced contributions can change rapidly.

In Fig.~\ref{fig:nonfwdG} we show the cross sections for the
$\omega$-Reggeon contribution alone and for the full amplitude
$T_{\rm Odd}+T_{P\otimes{\rm Odd}}+T_R$ with $C=1$ and $C=1.8$, using the
effective gluon masses $m_g=0.17$~GeV and $m_g=0.7$~GeV.  The value
$m_g=0.7$--$0.8$~GeV for perturbative gluons is reasonable~\cite{ZaP7} from
the point of view of the gluon correlation radius in the QCD vacuum,
$R_c\sim 0.27$~fm~\cite{ZaP8}.  The results for the pole form-factor
parametrisation are shown in the Appendix (Fig.~\ref{fig:nonfwdP}).

Photon exchange also contributes to the non-forward $C=-1$ amplitude,
through $\KL+p\to\KS+p$ and $\KL+e\to\KS+e$.  For $E_K\gtrsim 1$~TeV the
electron contribution is kinematically suppressed at
$q^2\gtrsim 0.2$~GeV$^2$, and for $m_g=0.17$~GeV the photon-exchange
modification of $d\sigma/dt$ is moderate ($\lesssim 10$--$15\%$) in the
region $q^2\gtrsim 0.2$~GeV$^2$ preferred below.  It can, however, become
important in the high-$|t|$ shoulder region, $q^2\gtrsim 1$~GeV$^2$, and in
the $m_g=0.7$~GeV scenario (in particular for $C\simeq 1.8$), where the
interpretation of an observed shoulder as an Odderon signal would be
ambiguous. The absorptive corrections in the quasi-eikonal model are qualitative in
nature; in addition, the competing photon-exchange amplitude must be kept
in mind whenever the measured $d\sigma/dt$ falls below
$\sim 10^{-6}$~mb/GeV$^2$ at $|t|\sim 1$~GeV$^2$.

\section{Possible experimental setup}
\label{sec:setup}
One may consider two main regeneration modes, namely the forward
coherent (elastic) mode, where the neutral kaons interacting with the
nuclei in the regenerator are deflected by unmeasurably small angles, and the
diffractive (non-forward) mode, where the neutral-kaon deflection is of the
order of several mrad.

Our considerations regarding possible experimental setups for measuring the
Odderon exchange via the regeneration $\KL\rightarrow\KS$ are based on the
existing CERN infrastructure.  The prerequisites are a source of \KL{}
mesons, a regenerator and a detector.  Let us consider the ATLAS interaction
point (IP) as the source of kaons from proton--proton collisions, although
it can be any major LHC detector if the conditions are favourable.  A
fraction of the \KL{} mesons which arrive at the regenerator is transformed
into \KS{} in the regenerator, and the aim is to measure the number of these
regenerated \KS{} mesons in a distant detector.  There is certainly a
variety of configurations, depending on the material and length of the
regenerator, on the distance of the regenerator from the ATLAS IP, and on
the distance of the detector from the regenerator.  While for the source of
kaons we choose the ATLAS IP, for the detector we consider the LHCf detector
type.  Currently, the LHCf experiment is placed in the TAN
absorber~\cite{hlTAN}, located 140~m from the ATLAS IP.  The LHCf experiment
has proved very useful in measuring the fluxes and energies of neutral
particles such as neutrons, photons and $\pi^0$ mesons.  In the LHCf
prospects for the special run in Run~3, there are plans to measure also
neutral kaons via their decays into two neutral pions, each of which then
decays into two photons.  The expected yield for this special run is of the
order of 1000~\Kz{} mesons with reliably measured fluxes, energies and decay
points.  One may propose to move the LHCf detector from the TAN absorber some
150~m downstream and to fill the aperture in the TAN with the regenerator
material.  This
would require drilling small apertures in the TAN body in front of and
behind the regenerator, such that the neutral kaons arriving from the ATLAS
IP are not absorbed in the TAN before they reach, and after they leave, the
regenerator.  The whole measurement would be performed only during a special
run of a duration similar to that in Run~3, and the two small apertures
would be re-filled after the end of this special run.  We therefore consider
the number of all measured neutral kaons to be at least 1000.  Another
important parameter of the experimental setup is the distance of the regenerator
from the ATLAS IP, 140~m.  We can also select the energy range of the
detected neutral kaons.

To this end, we need to consider two competing requirements.  On the one
hand, if the kaon energies are too high, many \KS{} mesons produced at the
ATLAS IP survive, owing to the Lorentz boost, up to the location of the
regenerator, and would contaminate the measurement with \KS{} mesons that do
not originate from regeneration (the statistical issues connected with this
contamination of the \KL{} beam by primary \KS{} mesons from the IP are
discussed separately in Section~\ref{sec:yield}).  This primary-\KS{}
background can be reduced by selecting low-energy kaons only.  However, as
discussed above, at low kaon energies the effect of the Odderon exchange on
the amplitude is small and hardly measurable, so high kaon energies are
required in the coherent forward regeneration case.  Thus, performing
coherent forward regeneration experiments at the LHC would require
a separate \KL{} beamline and non-trivial changes to the CERN tunnel
infrastructure, allowing a
$400$--$500$~m long free path for the \KL{} mesons before they reach the
regenerator.

As documented in Ref.~\cite{LHCf_photon}, the minimum photon energy measured
by LHCf is 200~GeV.  The photon energy spectrum reaches values of 6~TeV,
but, as follows directly from the survival probability
$P_S(L)=\exp[-L/(\gamma c\tau_S)]$ with $c\tau_S=2.68$~cm, the primary-\KS{}
contamination (i.e.\ the fraction of \KS{} reaching the regenerator at
$L=140$~m) is at the level of 27\% at $E_{K}\approx 2$~TeV
(where $\gamma c\tau_S\simeq 108$~m and hence
$P_S\simeq\exp(-140/108)\simeq 0.27$); it drops to 0.6\% at 0.5~TeV and to a
negligible level of about $2\times 10^{-6}$ at 0.2~TeV.  If we recall that
the \KS{} lifetime is about 570~times shorter than that of the \KL{} meson,
while its branching ratio into two neutral pions is 355~times larger, we
would have to restrict ourselves to rather low energies of the detected
neutral kaons in a fairly narrow energy range.  Even if the LHCf detector
and reconstruction techniques could be upgraded to measure at such low
energies, the Odderon effect would stay hidden and unmeasurable in the
coherent mode.  In the following subsection we examine the non-forward 
\KS{} regeneration mode.

\subsection{Regeneration in the non-forward mode}
\label{sec:nonfwd-setup}
As stated earlier, within the considered CERN infrastructure (ATLAS IP and
TAN absorber at 140~m from the IP) we have to restrict ourselves to a rather
narrow range
of low energies of the produced neutral kaons.  In this way we ensure that
the contamination of the \KL{} mesons arriving at the regenerator by
the primary \KS{} mesons is negligible.  This non-forward option would involve
hardware
changes not considered in the HL-LHC plans.  The first change is moving the
LHCf-type detector from the TAN absorber about 150~m downstream.  Since the
non-forward scattering of neutral kaons off the regenerator nuclei
deflects the kaons by small angles of several mrad, this LHCf-type detector
has to be placed further from the beam in the vertical direction.  We also
need to adapt the TAN absorber in such a way that the incoming and outgoing
neutral kaons are not blocked by the TAN material.  This would mean drilling
holes in front of and behind the regenerator, the latter at a small polar
angle, since the measured kaons will be deflected in that direction.  As the energy
spectrum of neutral kaons obtained from proton--proton collisions generated
with PYTHIA~8.3 shows, roughly half of the available kaon statistics can be
expected in the range 0.2--0.5~TeV, with a mean value around 280~GeV.  In
the accumulated sample, one would have to concentrate on a rather narrow
region around $|t|=0.5$~GeV$^2$ if the effective gluon mass is 0.17~GeV.
All the above requirements make this measurement statistically very
limited.

\section{Event-yield considerations}
\label{sec:yield}
A central limitation of an LHC-based regeneration measurement is the
contamination of the incoming \KL{} beam by primary \KS{} mesons produced at
the interaction point.  Since the regenerated signal is also observed
through $\KS\rightarrow\pi^0\pi^0$, even a very small surviving primary-\KS{}
component can generate a sizeable contribution to the observed decay-vertex
distribution and to its statistical uncertainty.

Let $L_{\rm IP}$ be the distance between the interaction point and the
regenerator.  If the initial neutral-kaon flux contains the ratio
$\varepsilon_S^{(0)}=\Phi_S(0)/\Phi_L(0)$, the corresponding ratio at the
regenerator entrance is
\begin{equation}
\label{eq:epsS}
\varepsilon_S(E_K,L_{\rm IP})
=\varepsilon_S^{(0)}
\exp\!\left[-\frac{L_{\rm IP}}{\gamma c\tau_S}
            +\frac{L_{\rm IP}}{\gamma c\tau_L}\right]
\simeq \varepsilon_S^{(0)}
\exp\!\left[-\frac{L_{\rm IP}}{\gamma c\tau_S}\right],
\end{equation}
where the last form uses $\tau_L\gg\tau_S$.  For equal \KS{} and \KL{}
production at the interaction point, $\varepsilon_S^{(0)}\simeq 1$.
Equation~(\ref{eq:epsS}) reproduces the survival fractions quoted in
Section~\ref{sec:setup}: for $L_{\rm IP}=140$~m the surviving primary-\KS{}
fraction is about 27\% at $E_K=2$~TeV, about $6\times 10^{-3}$ at
$E_K=0.5$~TeV, and negligible at $E_K=0.2$~TeV.  Requiring
$\varepsilon_S\leq\varepsilon_{\max}$ translates into
\begin{equation}
\label{eq:LIPmin}
L_{\rm IP}\;\geq\;\gamma c\tau_S\,\ln(1/\varepsilon_{\max})\,,
\end{equation}
i.e.\ $L_{\rm IP}\gtrsim 9.2\,(11.5)\,\gamma c\tau_S$ for
$\varepsilon_{\max}=10^{-4}\,(10^{-5})$, to be compared with the
$\gtrsim 30\,\gamma c\tau_S$ adopted in dedicated regeneration
experiments~\cite{KTeV_FNAL}, where \KS{} and \KL{} are produced in equal
proportion at the target (here $c\tau_S\gamma=\gamma\cdot 2.68$~cm).  For
the TAN location, $L_{\rm IP}=140$~m, this restricts the useful kaon
energies to $E_K\lesssim 0.3$~TeV, whereas the same purity at
$E_K\approx 2$~TeV would require $L_{\rm IP}\gtrsim 1$~km.  For comparison,
at an FPF-like distance, $L_{\rm IP}\simeq 620$--$650$~m~\cite{FPF}, the
estimate~(\ref{eq:epsS}) gives
\begin{equation}
\label{eq:FPF}
\exp\!\left[-\frac{L_{\rm IP}}{\gamma c\tau_S}\right]
\simeq (2\mbox{--}3)\times 10^{-3}\,,
\qquad E_K\simeq 2~{\rm TeV}\;\;(\gamma\simeq 4.0\times 10^{3})\,,
\end{equation}
consistent with the conclusion that coherent TeV-energy regeneration is
strongly disfavoured in a TAN/LHCf geometry by primary-\KS{} survival, but
remains conceptually viable in an FPF-like forward cavern, provided a
neutral-kaon beamline, a regenerator and a downstream
$\Kz\rightarrow\pi^0\pi^0$ detector can be implemented there.

The number of primary-\KS{} decays in a decay-vertex bin
$[X_i,X_i+\Delta X_i]$ downstream of the regenerator can be written as
\begin{equation}
\label{eq:NSprim}
N_i^{S,{\rm prim}}
=\Phi_K\,\varepsilon_S(E_K,L_{\rm IP})\,A_S\,
B(\KS\rightarrow\pi^0\pi^0)
\Big[e^{-X_i/\lambda_S}-e^{-(X_i+\Delta X_i)/\lambda_S}\Big]\,,
\end{equation}
where $\Phi_K$ is the \KL{} flux at the regenerator entrance,
$\lambda_S=\gamma c\tau_S$, and $A_S$ denotes the attenuation of the primary
\KS{} component in the regenerator material.  The corresponding CP-violating
$\KL\rightarrow\pi^0\pi^0$ contribution in the same bin is
\begin{equation}
\label{eq:NLCP}
N_i^{L,{\rm CP}}
=\Phi_K\,A_L\,B(\KL\rightarrow\pi^0\pi^0)
\Big[e^{-X_i/\lambda_L}-e^{-(X_i+\Delta X_i)/\lambda_L}\Big]\,,
\end{equation}
with $\lambda_L=\gamma c\tau_L$.  The useful quantity is not only the
surviving $\KS/\KL$ flux ratio at the regenerator, but the ratio of the
number of primary-$\KS\rightarrow\pi^0\pi^0$ decays to the number of
CP-violating $\KL\rightarrow\pi^0\pi^0$ decays in the same decay-vertex bin.
We therefore define the bin-wise contamination figure of merit
\begin{equation}
\label{eq:rSCP}
r_i^{S/{\rm CP}}\equiv\frac{N_i^{S,{\rm prim}}}{N_i^{L,{\rm CP}}}
=\varepsilon_S(E_K,L_{\rm IP})\,\frac{A_S}{A_L}\,
\frac{B(\KS\rightarrow\pi^0\pi^0)}{B(\KL\rightarrow\pi^0\pi^0)}\,
\frac{e^{-X_i/\lambda_S}-e^{-(X_i+\Delta X_i)/\lambda_S}}
     {e^{-X_i/\lambda_L}-e^{-(X_i+\Delta X_i)/\lambda_L}}\,.
\end{equation}
This ratio, rather than the survival probability alone, determines whether
the primary-\KS{} component can be subtracted with sufficient statistical
precision, and thus answers directly the practical question of how close the
regenerator may be placed for a given kaon-energy window.  Values
$r_i^{S/{\rm CP}}\ll 1$ correspond to a clean \KL{} reference sample, while
$r_i^{S/{\rm CP}}\gtrsim 1$ imply that the primary-\KS{} component dominates
the no-regenerator decay-vertex distribution and must be subtracted with a
correspondingly high precision.  Because
$B(\KS\rightarrow\pi^0\pi^0)\gg B(\KL\rightarrow\pi^0\pi^0)$ and
$\lambda_L\gg\lambda_S$ (in the first few \KS{} decay lengths the last
factor in Eq.~(\ref{eq:rSCP}) approaches $\lambda_L/\lambda_S\simeq 571$),
even $\varepsilon_S\sim 10^{-5}$--$10^{-4}$ can be relevant in the first few
\KS{} decay lengths behind the regenerator\footnote{Strictly speaking, the
surviving primary-\KS{} amplitude remains phase-coherent with the \KL{}
component, giving rise to the well-known vacuum-interference term
$\propto 2|\eta|\,e^{-L_{\rm IP}/2\lambda_S}
\cos(\Delta\omega_{LS}\tau+\phi_\eta)$.  Relative to the primary-\KS{}
intensity itself this term is suppressed by
$\sim 2|\eta|\,e^{+L_{\rm IP}/2\lambda_S}$ and can be included in the fit
together with the incoherent contributions (see, e.g.,
Refs.~\cite{Carosi,E773} for measurements of the vacuum-interference
phase); a dedicated coherent treatment
would be required for a precision extraction.}.
In our case we therefore aim at a $10^{-5}$--$10^{-4}$ contamination of the
\KL{} flux by primary \KS{} mesons from the IP and expect the ratio of the
fitted primary-\KS{} to the \KL{} reference yields to stay below
$\approx 0.2$, so that the background due to primary \KS{} decays can be
reliably subtracted.

For the statistical analysis we define the measured ratio in a decay-vertex
bin as
\begin{equation}
\label{eq:Ri}
R_i\equiv
\frac{N_i^{\rm on}-B_i^{S,{\rm on}}-B_i^{n,{\rm on}}}
     {N_i^{\rm off}-B_i^{S,{\rm off}}}\,,
\end{equation}
where ``on'' and ``off'' denote running with and without the regenerator,
$B_i^{S}$ is the primary-\KS{} component estimated from
Eqs.~(\ref{eq:epsS})--(\ref{eq:rSCP}) and constrained by the fit of the
exponentially falling \KS{} term in the no-regenerator decay-vertex
distribution, and $B_i^{n}$ is the neutron-induced secondary-\KS{}
background produced in the regenerator (Section~\ref{sec:backgrounds}).  Its
variance should be obtained from
\begin{equation}
\label{eq:dRi}
(\delta R_i)^2=\sum_{x_j}
\left(\frac{\partial R_i}{\partial x_j}\right)^2 (\delta x_j)^2
+2\sum_{j<k}\frac{\partial R_i}{\partial x_j}
             \frac{\partial R_i}{\partial x_k}\,{\rm Cov}(x_j,x_k)\,,
\end{equation}
with $x_j=\{N_i^{\rm on},N_i^{\rm off},B_i^{S,{\rm on}},B_i^{S,{\rm off}},
B_i^{n,{\rm on}}\}$ and with additional nuisance parameters for the
attenuation, efficiency and flux normalisation in a full detector-level
likelihood analysis.  Note that the primary-\KS{} contamination not only
biases $R_i$ if unsubtracted, but also inflates the statistical uncertainty
of the measured ratio through the (anti)correlated subtraction terms in
Eq.~(\ref{eq:dRi}); the accepted geometry and energy range should therefore
keep the fitted primary-\KS{} component small enough that its statistical
and systematic uncertainties remain subdominant to the Odderon-induced
change of the regenerated decay-vertex distribution.

\section{Suppression of specific backgrounds}
\label{sec:backgrounds}
\subsection{Neutron-induced background}
\label{sec:nbkg}
Neutral-kaon beams are always accompanied by a neutron flux, which
contaminates the regenerated \KS{} sample with \KS{} mesons created inside
the regenerator material by inelastic neutron--nucleus (or neutron--proton)
interactions; the forward neutron flux itself is well constrained by the
LHCf measurements~\cite{LHCfNeutron}.  The vast majority of such
neutron-induced \KS{} mesons is accompanied by several charged particles,
which allows one to efficiently veto (filter out) such \KS{} mesons using an
active regenerator~\cite{ActiveRegenerator}.  However, a small fraction of
newly created (not regenerated) \KS{} mesons can emerge from (semi)exclusive
processes that are not accompanied by detectable charged particles.  Using
the UrQMD~\cite{UrQMD} simulation, we have identified two major types of
inelastic QCD reactions which do not involve energetic charged particles in
the final state (assuming a slow nucleus remnant $A'$):
\begin{equation}
n(udd)+A \longrightarrow \Sigma^0/\Lambda^0(uds)+\Kz(d\bar s)+A'
\label{eq:nLamK}
\end{equation}
and
\begin{equation}
n(udd)+A \longrightarrow n'(udd)+\Kzb(\bar d s)+\Kz(d\bar s)+A'\,,
\label{eq:nKK}
\end{equation}
where $A$ is a nucleus of the regenerator (or just a proton in the case of
LH$_2$).  Both reactions~(\ref{eq:nLamK}) and~(\ref{eq:nKK}) can also be
accompanied by $\pi^0$ or $\eta$ mesons decaying into the neutral
$\gamma\gamma$ state without any charged particles in the final state.  If
the momentum of the {\bf elastically} scattered nucleus $A'$ is too small to
be detected by the active-regenerator setup, the above two reactions
represent a neutron-induced background to the regenerated \KS{} mesons which
cannot be filtered out by the active regenerator~\cite{ActiveRegenerator}.

A precise Geant simulation, taking into account all the material surrounding
the \KL{} beam pipe and the detector, would be needed to estimate the
contamination of the regenerated $\KL\rightarrow\KS$ sample by the
neutron-induced \KS{} mesons.  We do not attempt such detailed simulations
here, since the results depend on the specific experimental setup.  Instead,
we consider the possibility of subtracting the $n$-induced \KS{} background
using the double-regenerator technique employed, e.g., by the E773
experiment~\cite{E773}.

\begin{figure}[t!]
\begin{center}
\includegraphics[width=0.42\textwidth]{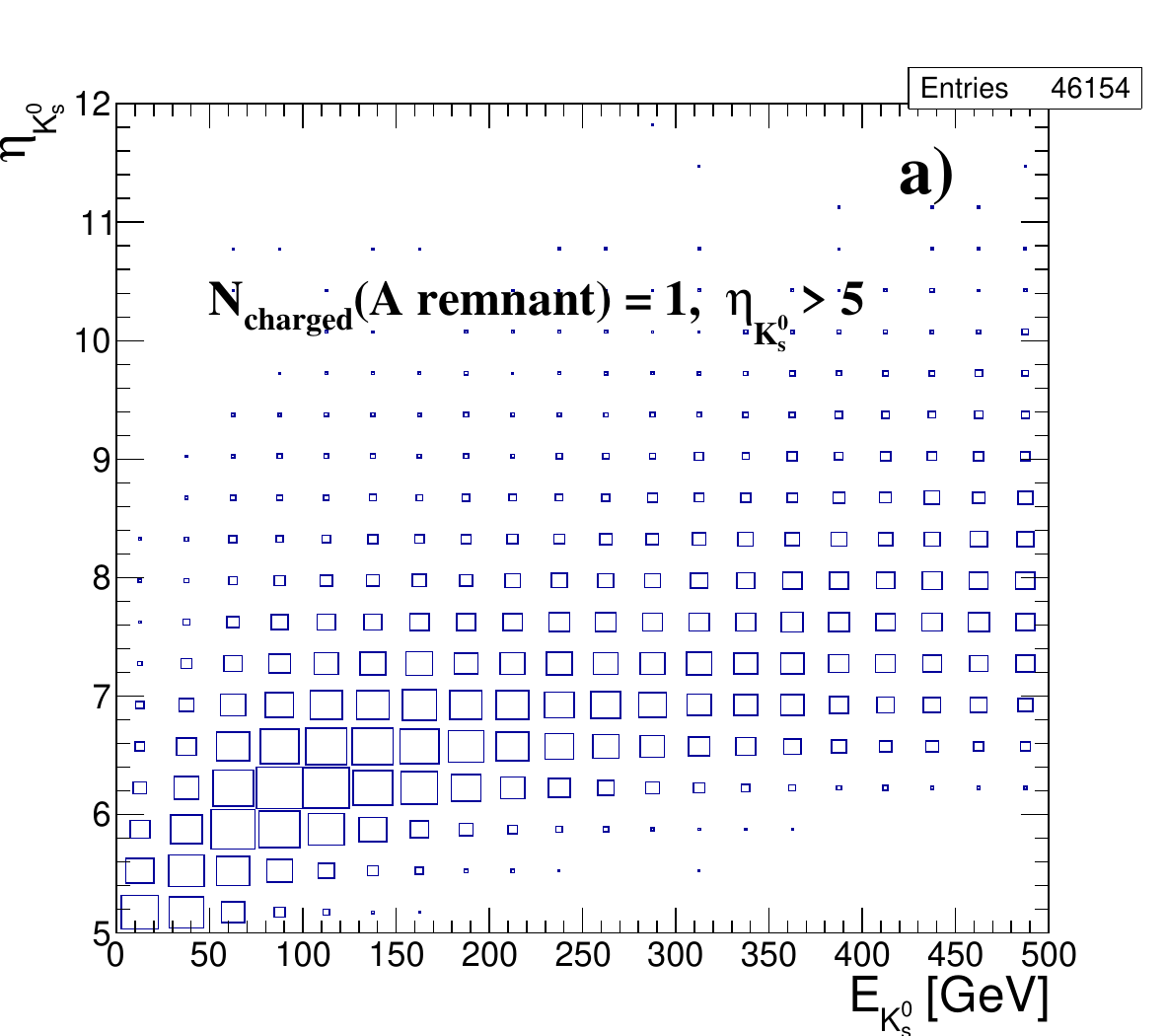}
\includegraphics[width=0.42\textwidth]{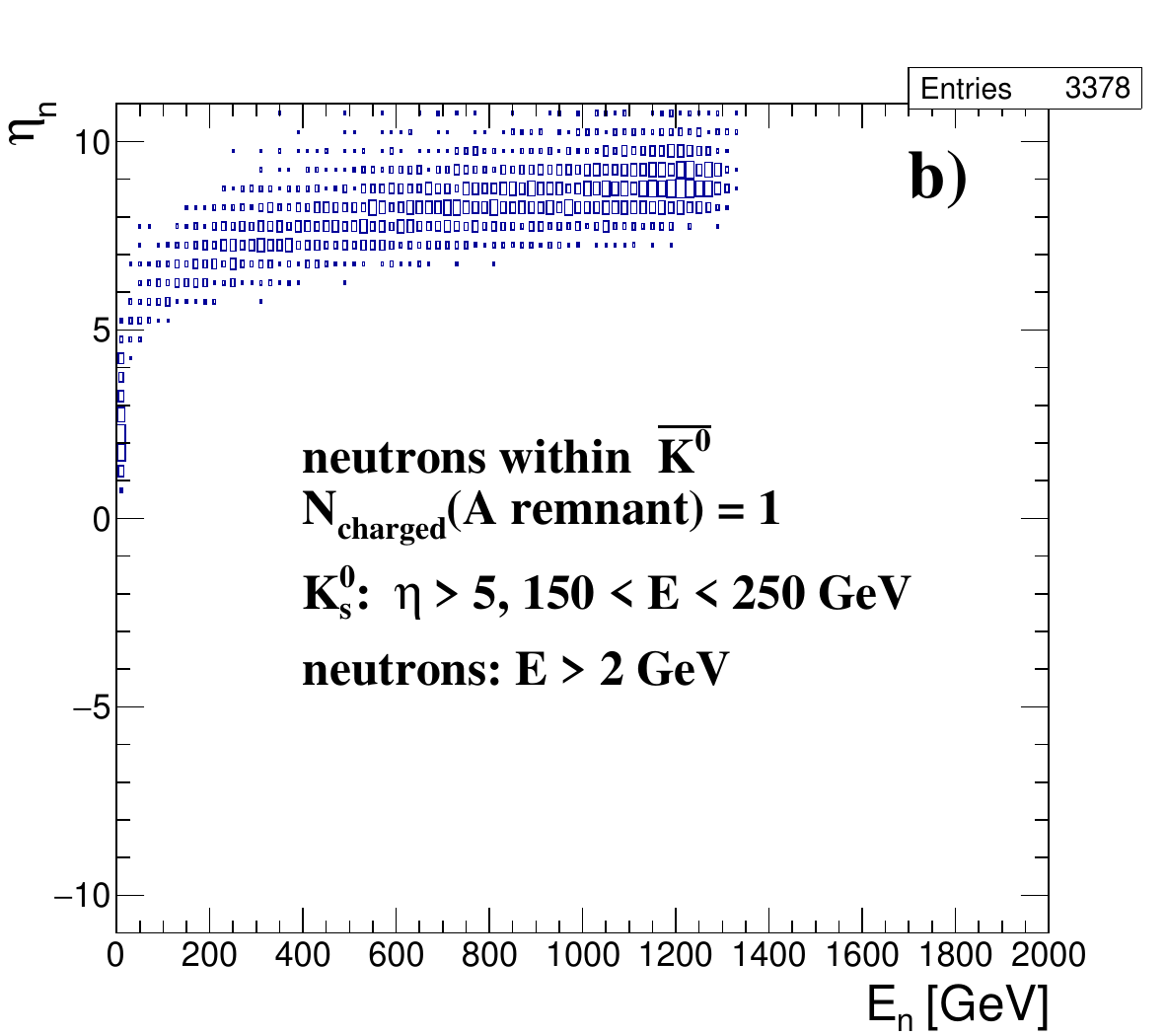}\\
\includegraphics[width=0.42\textwidth]{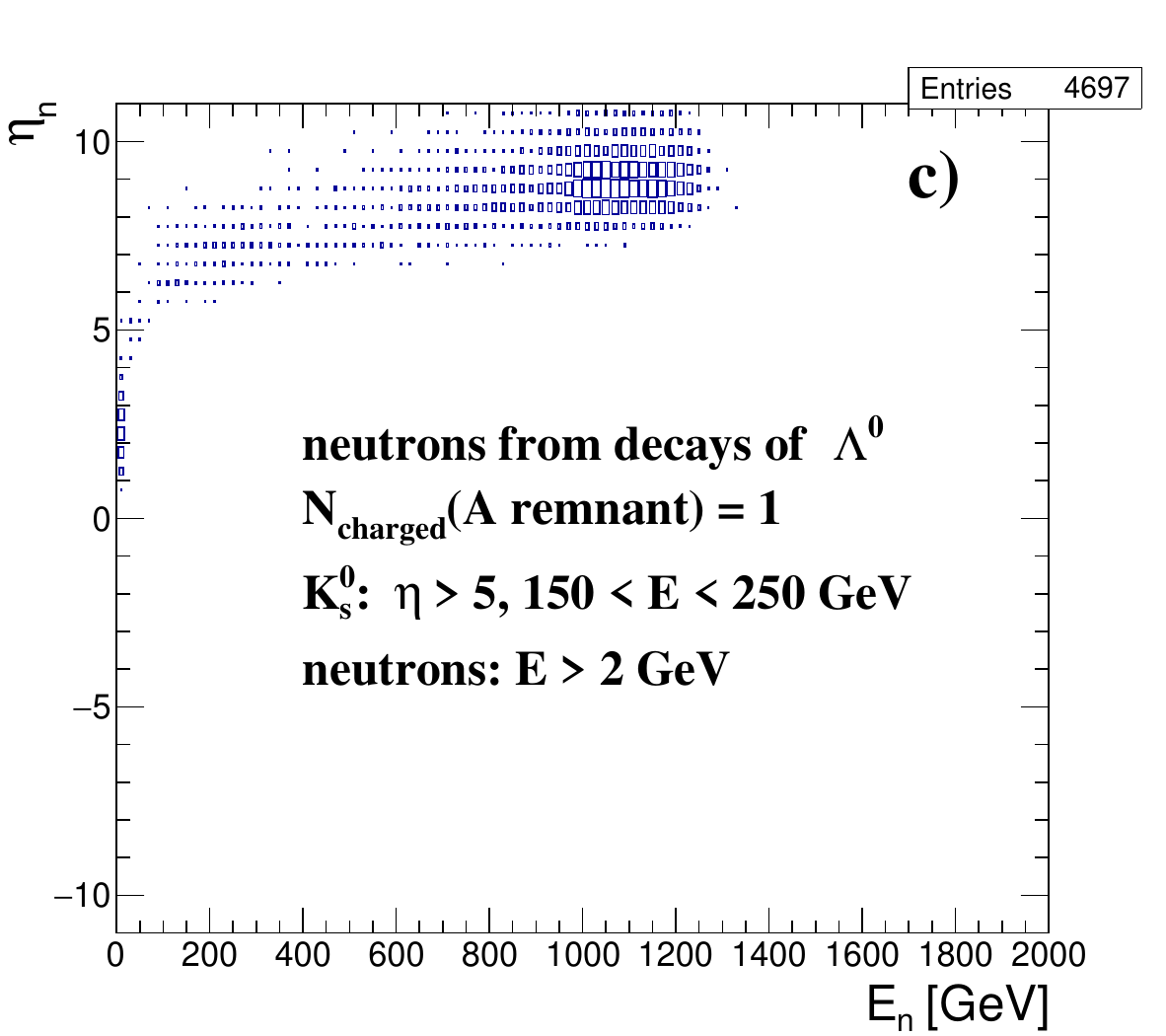}
\end{center}
\caption{PYTHIA~8.3 particle-level information from $n$+Pb reactions in the
fixed-target mode with a neutron energy of 1.5~TeV.  Shown are
two-dimensional distributions of energy versus pseudorapidity after
requiring no charged particles (except the nucleus remnant) and
$\eta_{\KS}>5$.  In a) the information on the \KS{} mesons is shown, while
in b) the neutrons accompanying both \Kz{} mesons in the
reaction~(\ref{eq:nKK}) and in c) the neutrons from decays of $\Lambda^0$ in
the reaction~(\ref{eq:nLamK}) are shown, after additionally requiring
$150<E_{\KS}<250$~GeV and $E_{\rm neutron}>2$~GeV.}
\label{fig:PY8.3}
\end{figure}

To estimate quantitatively the probability of the $n$-induced background
reactions~(\ref{eq:nLamK}) and~(\ref{eq:nKK}), we have used the
particle-level information from the PYTHIA~8.3 generator~\cite{PYTHIA8.3}.
These simulations should be regarded as order-of-magnitude diagnostics of
the dangerous neutral-only topologies rather than as precision predictions
for exclusive forward strangeness production; a final background estimate
would require a dedicated detector-level simulation, preferably
cross-checked with other hadronic generators, with the dedicated
forward-physics PYTHIA tune~\cite{FwdTune}, and with target data.  The
reactions of neutrons with neutrons, protons and with d, C and Pb nuclei in
the fixed-target configuration have been generated using the {\it Angantyr}
model, with the neutron energy initially fixed at 1.5~TeV.
High-statistics samples ($10^9$ events each) allowed us to study the energy
and charged-particle flow in the forward direction for the two considered
(semi)exclusive final states~(\ref{eq:nLamK}) and~(\ref{eq:nKK}), which is
essential for separating this type of \KS{} background events from the
$\KL\rightarrow\KS$ regeneration signal.  Figure~\ref{fig:PY8.3} shows
two-dimensional distributions of energy versus pseudorapidity of the \KS{}
mesons (in~a)) and of the neutrons (in~b) and~c)) appearing in the final
state when energetic neutrons hit nuclei in the regenerator (Pb as an
example) and no charged particles are present in the final state except the
nucleus remnant.  Plots~b) and~c) show the neutron flows for the
(semi)exclusive reactions~(\ref{eq:nLamK}) and~(\ref{eq:nKK}) for events
faking the regenerated \KS{} signal in the LHCf acceptance region
($\eta_{\KS}>5$ and $150<E_{\KS}<250$~GeV).

Table~\ref{tab:nA} shows the inelastic production cross sections together
with the fiducial cross sections obtained after applying all the cuts
imposed on the \KS{} mesons and neutrons described in the caption of
Fig.~\ref{fig:PY8.3}.
\begin{table}[t!]
\begin{center}
\parbox{14.6cm}{\caption{The inelastic cross sections,
$\sigma_{\rm inel}$, for the interaction of a high-energy neutron with a
fixed-target {\it neutron}, proton and nuclei of type deuteron, C and Pb,
and the fiducial cross sections,
$\sigma_{\rm fid}=\sigma_{\rm inel}N_{\rm surv}/N_{\rm gen}$, for selecting
exclusive events with the charge-less final states of
types~(\ref{eq:nLamK}) and~(\ref{eq:nKK}), shown in the third and fourth
rows, obtained after applying the cuts described in the caption of
Fig.~\ref{fig:PY8.3}~b) and~c).  Only statistical uncertainties are
quoted.}\label{tab:nA}}\vspace*{0.3cm}
\begin{tabular}{|c|c|c|c|c|c|}
  \hline
Reaction $n+A$ & $A$=neutron & $A$=proton & $A$=deuteron & $A$=$^{12}$C & $A$=$^{208}$Pb \\
  \hline
$\rightarrow X, \ \ \ \ \ \ \sigma_{\rm inel}$ [mb] & 36 & 36 & 65 & 256 & 1640 \\
  \hline
$\rightarrow \Lambda^0\!+\!K^0,\ \sigma_{\rm fid}$ [$\mu$b] & 0.84 $\pm$ 0.01 & 0.82 $\pm$ 0.01 & 1.19 $\pm$ 0.01 & 1.90 $\pm$ 0.02 & 7.70 $\pm$ 0.11\\
  \hline
$\rightarrow \bar K^0\!+\!K^0,\ \sigma_{\rm fid}$ [$\mu$b] & 0.65 $\pm$ 0.01 & 0.62 $\pm$ 0.01 & 0.92 $\pm$ 0.01 & 1.40 $\pm$ 0.02 & 5.54 $\pm$ 0.10 \\
  \hline
\end{tabular}
\end{center}
\end{table}

Two observations follow from Table~\ref{tab:nA}.  First, the small differences between the
neutron- and proton-target fiducial cross sections cannot be explained by
purely statistical uncertainties; investigating their origin goes beyond
the scope of this study.
Second, as the mass of the regenerator nucleus increases, fewer and fewer
events survive the signal-selection cuts, but this is overcompensated by the
faster increase of the inelastic $nA$ cross section, so that the fiducial
background cross section per nucleus grows monotonically with $A$.  Hence
the interaction of neutrons with heavier nuclei represents a very dangerous
background.  Although most of the energetic neutrons going very forward (see
Fig.~\ref{fig:PY8.3}~b) and~c)) can be expected to be identified in a
ZDC-type detector, the small fraction of neutrons with energies below a ZDC
acceptance threshold (e.g.\ 50~GeV) still represents a dangerous background,
owing to the huge inelastic $nA$ cross section.  A detailed look at the
event records of these remaining (low neutron-energy) background events
shows that the slow neutrons are always accompanied by energetic and very
forward-going neutral pions or $\eta$ mesons (conserving the longitudinal
momentum of the incoming neutron).  The photons from the forward-going $\pi^0$ and $\eta$ decays can
in principle also be detected, e.g.\ in an additional electromagnetic
veto calorimeter.

\begin{figure}[t!]
\begin{center}
\includegraphics[width=0.8\textwidth]{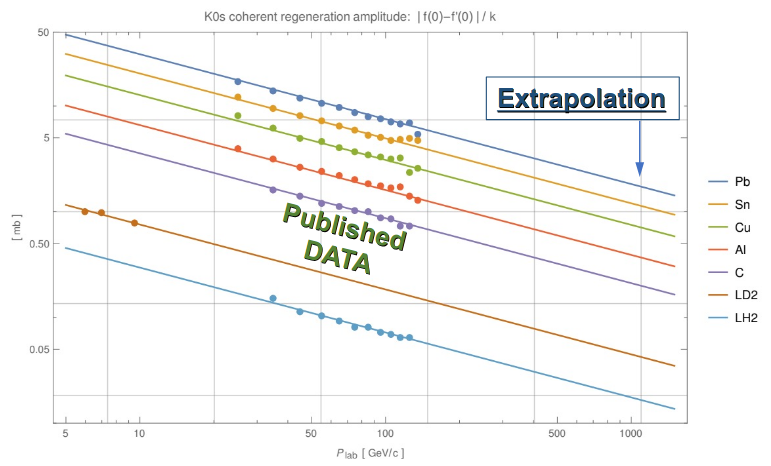}
\parbox{13.5cm}{\caption{Momentum dependence of the coherent regeneration
amplitude $|f(0)-\bar f(0)|/k$ for various regenerator materials,
extrapolated (straight lines) from the measured data compiled from
Refs.~\cite{Aronson-FNAL,Bock,Gsponer,Roehrig,Molzon}.}
\label{fig:RegAMP}}
\end{center}
\end{figure}

\begin{figure}[t!]
\begin{center}
\includegraphics[width=0.62\textwidth]{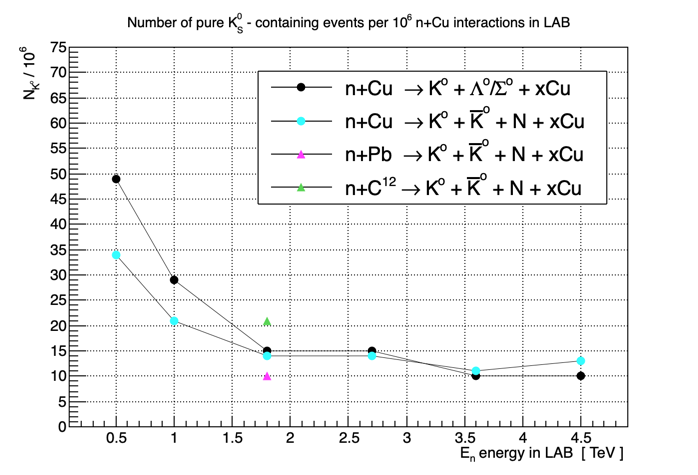}
\parbox{13.5cm}{\caption{UrQMD estimate of the yields of the (semi)exclusive
neutral-only reactions~(\ref{eq:nLamK}) and~(\ref{eq:nKK}) per $10^6$
inelastic $n$+Cu interactions, as a function of the neutron energy in the
laboratory frame.  The single points for $n$+Pb and $n+^{12}$C at
$E_n=1.8$~TeV are shown for comparison.  Both channels {\it decrease} with
increasing neutron energy.}\label{fig:UrQMDenergy}}
\end{center}
\end{figure}

From Fig.~\ref{fig:RegAMP} one observes that the regeneration amplitude
increases with the number of nucleons in the regenerator nuclei.  For
example, at $P_{\rm lab}=100$~GeV/$c$ the regeneration power of carbon is
$\approx 10$~times smaller than that of lead, and this ratio (1:10) of the
regeneration amplitudes may remain constant at higher $P_{\rm lab}$ momenta
(here the same linear log--log slope is assumed for all materials).

The double-regenerator idea can be written transparently as
\begin{equation}
\label{eq:double}
N_i=S_i+B_i\,,\qquad
S_i=\Phi_K\,\epsilon_i\,P_i^{\rm reg}\,,\qquad
B_i=\Phi_n\,\epsilon_i^{\,n}\,n_iL_i\,\sigma^{\,nA_i}_{\rm bg}\,,
\qquad i=1,2\,,
\end{equation}
where $S_i$ is the genuine regenerated signal and $B_i$ the neutron-induced
neutral-only background for the regenerator configuration (material and
length) $i$; $\Phi_K$ and $\Phi_n$ denote the integrated
\KL{} and neutron fluxes impinging on the regenerator, $\epsilon_i$
($\epsilon_i^{\,n}$) the detection efficiencies for the regenerated
(neutron-induced) \KS{} mesons, $P_i^{\rm reg}$ the regeneration
probability, $n_iL_i$ the number of target nuclei per unit area, and
$\sigma^{nA_i}_{\rm bg}$ the fiducial background cross sections of
Table~\ref{tab:nA}.  If the two regenerator configurations are chosen such
that the neutron-induced backgrounds are equal, $B_1=B_2$ (i.e.\
$n_1L_1\sigma^{nA_1}_{\rm bg}=n_2L_2\sigma^{nA_2}_{\rm bg}$), while the
regeneration probabilities differ, $P_1^{\rm reg}\neq P_2^{\rm reg}$, then
the measured {\it difference} of the \KS{} yields,
$N_1-N_2=S_1-S_2$, is free of the neutron-induced background and directly
proportional to the difference of the regeneration signals.  Conversely, a
pair with equal regeneration probability but different backgrounds
determines the background normalisation $\Phi_n\epsilon^{\,n}$ from
the data themselves, so that the background-free combination
$S=(\tilde\sigma_1 N_2-\tilde\sigma_2 N_1)/(\tilde\sigma_1-\tilde\sigma_2)$,
with $\tilde\sigma_i\equiv n_iL_i\,\sigma^{nA_i}_{\rm bg}$, can be formed.
Provided the regeneration-amplitude ratio for Pb and C is
significantly different from the ratio of the \KS{}-background production
cross sections of the (semi)exclusive reactions~(\ref{eq:nLamK})
and~(\ref{eq:nKK}) on Pb and C -- and Table~\ref{tab:nA} together with
Fig.~\ref{fig:RegAMP} indicates that it is (a factor $\approx 10$ in
amplitude versus $\approx 4$ in the background cross section per
nucleus) -- such non-degenerate pairs of active regenerators can indeed be
constructed.

The above considerations are significantly simplified, and the cross
sections of the reactions~(\ref{eq:nLamK}) and~(\ref{eq:nKK}) may vary
significantly with the neutron energy.  In fact, our UrQMD simulations
indicate {\it decreasing} yields of the (semi)exclusive chargeless
reactions~(\ref{eq:nLamK}) and~(\ref{eq:nKK}) with increasing neutron
energy, see Fig.~\ref{fig:UrQMDenergy}.  At the same time, the
two-regenerator considerations rely on an extrapolation of the {\it
coherent} regeneration amplitudes (Fig.~\ref{fig:RegAMP}) into an
experimentally unverified momentum region, assuming that the theoretical
justification~\cite{GilmanREG} for the equal slopes of the $P_{\rm lab}$
dependence (for different nuclei) remains valid also in the {\it
non-forward} regeneration case.

It is clear that the method described above (if indeed applicable) would
require a detailed measurement of the neutron-induced (semi)exclusive \Kz{}
production on various targets, which may turn out to be an interesting
research topic on its own (possibly related to QCD string-breaking
mechanisms involving strange quarks).  Indeed, the recently established
isospin violation in kaon production in nucleus--nucleus
collisions~\cite{NA41isoVIOLA} may be interpreted as an unexpectedly
{\it asymmetric} breaking of the QCD string into $(d\bar d)$ and
$(u\bar u)$ quark pairs~\cite{GermanEXPL}.  For a clarification of this
$u/d$ isospin asymmetry~\cite{NA41isoVIOLA}, another (semi)exclusive
reaction could be studied, namely
\begin{equation}
n(udd)+A\longrightarrow K^+(u\bar s)+K^-(\bar u s)+n'(udd)+A'\,,
\label{eq:nKpKm}
\end{equation}
which differs from the reaction~(\ref{eq:nKK}) just by the substitution of
the $(d\bar d)$ pair by a $(u\bar u)$ pair.  A recent theoretical
analysis~\cite{GermanEXPL} suggests that the probability ratio can be
$P(u\bar u):P(d\bar d)\approx 3:1$.

\subsection{Triple-Regge background}
\label{sec:tripleR}
There is another potential background source caused by the inelastic
$\omega$-Reggeon exchange in a triple-Regge topology~\cite{Mueller,KKPT1,KKPT2}.  In
comparison with the elastic cross section $d\sigma_{\rm el}/dt$, which is
proportional to
$$\beta^2_N\beta^2_K\,(s/s_0)^{2(\alpha_\omega-1)}$$
(here $s_0=1$~GeV$^2$, and $\beta_N$, $\beta_K$ are the couplings of the
$\omega$-Reggeon to the nucleon and to the kaon), the triple-Regge
contribution $M^2_X\,d\sigma/dt\,dM^2_X$ contains the factors
$$\beta_{{\rm Pom}\mbox{-}N}\,\beta_{(\omega\omega\mbox{-}{\rm Pom})}\,
\beta^2_K\,(M^2_X/s_0)^{\alpha_{\rm Pom}-1}\,(1-x)^{2(1-\alpha_\omega)}\,.$$
That is, to get the triple-Regge contribution (assuming
$\alpha_{\rm Pom}=1$ and $\alpha_\omega=0.5$), we have to multiply the
elastic result by
$$\frac{\beta_{(\omega\omega\mbox{-}{\rm Pom})}\,\beta_{\rm Pom}}
{\beta_N^2}\,\frac{s\,(1-x)}{s_0}\,,$$
with $x=M^2_X/s$; for a typical inelastic event we expect $x\sim 1/4$.
Although the vertex $\beta_{(\omega\omega\mbox{-}{\rm Pom})}$ is currently
not known, we use, for an order-of-magnitude estimate, the assumption
$\beta_{{\rm Pom}\mbox{-}N}\,\beta_{(\omega\omega\mbox{-}{\rm Pom})}
=\beta^2_N$.  That is, we expect
$$\frac{M^2_X\,d\sigma}{dt\,dM^2_X}\sim
\frac{d\sigma_{\rm el}}{dt}\,\frac{s\,(1-x)}{s_0}\,.$$

According to Figs.~\ref{fig:nonfwdG} and~\ref{fig:nonfwdP}, in the
non-forward (finite-$t$) regeneration on protons the $\omega$
contribution becomes comparable to that of the Odderon at
$p_{\rm lab}\simeq 200$~GeV, while for the forward ($t\to 0$) amplitude
entering the coherent regeneration
(see Fig.~4 of Ref.~\cite{OddKiev}) this happens only at
$p_{\rm lab}\sim 2$~TeV.  Since the regenerator nuclei dissociate in the
triple-Reggeon exchange, this background should be reducible by vetoing
forward-going charged particles, in the same way as for the neutron-induced
background; the residual neutral-dissociation channels should, however, be
included in the same detector-level background study as the neutron-induced
topologies.

\section{Benchmark signal and background rates}
\label{sec:benchmarks}
It is useful to summarise the above analysis in terms of explicit signal and
background scales.

{\it Non-forward mode.}  From Fig.~\ref{fig:nonfwdG} ($m_g=0.17$~GeV,
Gaussian form factors), the Odderon-dominated window
$0.4\lesssim|t|\lesssim 1$~GeV$^2$ at $p_K=0.2$--$0.8$~TeV corresponds to a
per-nucleon regeneration cross section
$\sigma_{\rm sig}\approx 0.05$--$0.1~\mu$b, exceeding the $\omega$-Reggeon
contribution there by a factor of $3$--$10$ at 200~GeV and by orders of
magnitude at 800~GeV (with a factor of a few model uncertainty from the
form-factor choice and from $C$; for $m_g=0.7$~GeV the signal drops by more
than two orders of magnitude and moves to $|t|\gtrsim 1$~GeV$^2$).  The
dangerous neutral-only neutron-induced background, summing the two channels
of Table~\ref{tab:nA}, amounts to $\sigma_{\rm bg}\approx 1.5~\mu$b per
nucleon (about 3.3~$\mu$b per C and 13.2~$\mu$b per Pb nucleus).  Hence,
before any active suppression,
\begin{equation}
\label{eq:BoverS}
\frac{B}{S}\;\approx\;20\,r_n\,,\qquad
r_n\equiv\frac{\Phi_n}{\Phi_{K}}\Big|_{\rm accepted\ window}\,,
\end{equation}
where $r_n$ is the neutron-to-\KL{} flux ratio at the regenerator in the
accepted energy window; the forward neutron flux itself is constrained by
the LHCf data~\cite{LHCfNeutron}, while $r_n$ for a specific installation
must be obtained from a dedicated simulation of the actual geometry.
For $r_n=\mathcal{O}(1$--$10)$, the neutral-only background must therefore
be suppressed by one to two orders of magnitude (a factor of
$\approx\!20$--$200$) to reach $S/B\gtrsim 1$;
this is precisely the suppression that the combination of the
active-regenerator charged-particle veto, the forward-neutron (ZDC-type) and
forward-photon tagging, and the double-regenerator subtraction of
Section~\ref{sec:nbkg} is expected to provide.  
We stress that Eq.~(\ref{eq:BoverS}) is a per-nucleon (hydrogen-target)
estimate: for C and Pb regenerators the neutral-only background grows to
about 3.3 and 13.2~$\mu$b per nucleus (Table~\ref{tab:nA}), but the
incoherent-regeneration signal grows simultaneously with the effective
number of quasi-free nucleons, so that the per-nucleus signal-to-background
ratio requires the nuclear (incoherent) regeneration cross section, which
we leave to a dedicated study.

{\it Coherent forward mode.}  For coherent regeneration at
$E_K\approx 2$~TeV, Fig.~\ref{fig:Our_prediction} shows that the
conservative benchmark $|\dphio|=20^{\circ}$ modifies the
$\Kz\rightarrow\pi^0\pi^0$ intensity by $20$--$40\%$ at
$20\lesssim\Delta X\lesssim 150$~m behind the Cu, C or Pb regenerators.
With $N\approx 10^3$ reconstructed $\Kz\rightarrow\pi^0\pi^0$ decays (the
scale of the anticipated LHCf special-run yield) distributed over several
decay-vertex bins, the statistical precision per bin is $15$--$20\%$, i.e.\
the Odderon-induced distortion would appear at the $(1$--$2)\sigma$ level
per bin and at the $\approx\!3\sigma$ level overall; a $\geq\!5\sigma$
demonstration would require $\mathcal{O}(10^4)$ reconstructed decays.
Moreover, at $L_{\rm IP}=140$~m the primary-\KS{} survival of $\approx 27\%$
precludes the measurement altogether, and even at an FPF-like
$L_{\rm IP}\approx 620$--$650$~m the surviving fraction of
$(2$--$3)\times 10^{-3}$, amplified by the branching-ratio and lifetime
factors in Eq.~(\ref{eq:rSCP}), leaves the primary-\KS{} component dominant
in the first \KS{} decay lengths and requires the subtraction procedure of
Section~\ref{sec:yield}.  Finally, the electromagnetic $C=-1$ amplitude of
Section~\ref{sec:coherent} constitutes an irreducible theory
component of the systematic uncertainty, at the level of the
Pomeron$\otimes$Odderon signal itself for
$m_g=0.17$~GeV (and dominant for $m_g=0.7$~GeV), unless constrained, e.g.,
by measurements on targets with different $Z/A$.

\section{Summary}
\label{sec:summary}
We have reconsidered the feasibility of using high-energy $\KL\to\KS$
regeneration as an independent probe of the crossing-odd (Odderon) exchange
at the LHC, with the \KL{} mesons produced, e.g., at the ATLAS interaction
point.  The main novelty of the study is the recasting of the original
Odderon-regeneration idea~\cite{OddKiev,OddZach} into the specific
constraints of the LHC infrastructure around the ATLAS interaction region.
From the practical point of view, the central new element is the
non-forward regeneration mode; in addition, we quantify the absorption in
realistic regenerators, the primary-\KS{} contamination, the competing
electromagnetic $C=-1$ amplitude, and the neutron-induced backgrounds.

Kinematic constraints for the coherent forward regeneration of 1--2~TeV
kaons and for the non-forward (sometimes called diffractive) regeneration of
$\approx\!0.2$--$0.8$~TeV kaons at the LHC have been quantified.  The
location of thick metallic (Cu, Pb) or carbon regenerators at (or behind)
the TAN absorber~\cite{hlTAN}, with an LHCf-type calorimeter~\cite{LHCf}
observing the $\pi^0\pi^0\to 4\gamma$ decay vertices, has been considered.

For the coherent regeneration of 1--2~TeV kaons, a conservative
Odderon-phase benchmark of $|\dphio|=20^{\circ}$ produces a $20$--$40\%$
distortion of the decay-vertex distribution 20--150~m behind the
regenerator, so that $\mathcal{O}(10^3)$ reconstructed
$\Kz\to\pi^0\pi^0$ decays would yield an effect only at the
$\approx\!3\sigma$ level and
$\mathcal{O}(10^4)$ decays are needed for a definitive observation.  The
present TAN/LHCf baseline ($L_{\rm IP}=140$~m) is excluded for this purpose
by the $\approx\!27\%$ primary-\KS{} contamination at 2~TeV, whereas an FPF-like
distance of about $620$--$650$~m suppresses the primary-\KS{} component to
the few-$10^{-3}$ level; a non-trivial modification of the CERN tunnel
infrastructure (a neutral-kaon beamline with a regenerator and a downstream
$\pi^0\pi^0$ detector) would be required, which may be relevant for a Long
Shutdown period or for the planned FPF facility~\cite{FPF}.  In addition,
the photon-exchange $C=-1$ amplitude is comparable to the
Pomeron$\otimes$Odderon contribution in the coherent mode and must be
constrained -- for instance, by comparing targets with different $Z/A$ --
before the coherent mode can be interpreted as an Odderon measurement.

Concerning the non-forward \KS{} regeneration at $0.2$--$0.8$~TeV, the
existing CERN infrastructure may provide an opportunity for such an
experiment: the Odderon-induced cross section in the window
$0.4\lesssim|t|\lesssim 1$~GeV$^2$ is expected at the level of
$0.05$--$0.1~\mu$b per nucleon for $m_g=0.17$~GeV, well above the
$\omega$-Reggeon contribution.  The measurement will, however, be contaminated by
severe neutron-induced and triple-Reggeon backgrounds: the neutral-only
neutron-induced \KS{} production alone amounts to $\approx\!1.5~\mu$b per
nucleon, so that a background suppression by one to two orders of
magnitude (depending on the neutron-to-kaon flux ratio) is required.  For
this purpose a specific active double-regenerator setup would be needed,
supplemented by a detailed understanding of the (semi)exclusive chargeless
QCD reactions yielding almost pure $\Sigma^0/\Lambda^0+\Kz$ or
$\Kz+\Kzb$ final states.  In its present form the study therefore
establishes the relevant signal and background scales and identifies the
measurements needed for a realistic sensitivity projection.  A definitive
proposal would require a detector-level simulation and a quantitative fit
connecting the benchmark Odderon phase shift to modern constraints on the
crossing-odd amplitude.

\section*{Acknowledgements}
Computational resources were provided by the e-INFRA CZ project
(ID:90254), supported by the Ministry of Education, Youth and Sports
(MEYS) of the Czech Republic. M.T.~is supported by the MEYS of the Czech
Republic within the project LM2023040 and P.F.~by project CZ.02.01.01/00/22\_008/0004632. V.A.K.~thanks the Institute of Physics of the Czech Academy of Sciences, Prague for hospitality.


\appendix
\section{Supplementary non-forward-regeneration predictions}
\label{app:supp}
Here we show our predictions for the $d\sigma/dt$ cross section of the
non-forward \KS{} regeneration with and without the Odderon exchange for the
pole form factors, and the underlying Odderon amplitudes $T_{\rm Odd}$ and
$T_{\rm Odd}+T_{P\otimes{\rm Odd}}$.

\begin{figure}[h!]
\begin{center}
\includegraphics[width=0.72\textwidth]{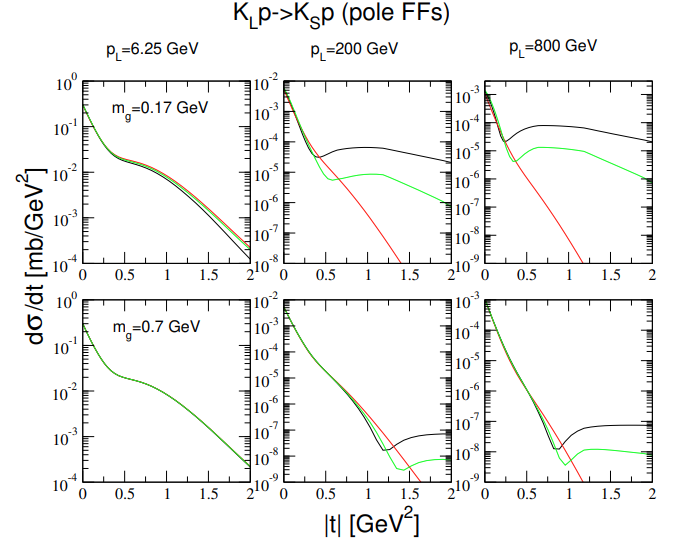}
\caption{Predictions of the cross section (as a function of $|t|$) for the
non-forward regeneration of \KS{} on protons, for three values of the kaon
momentum and two values of the effective gluon mass, using the pole form
factors.  The red curve shows the cross section due to the
$\omega$-Reggeon contribution alone, and the black and green curves show the
predictions for the total amplitude
$(T_{\rm Odd}+T_{P\otimes{\rm Odd}}+T_R)$ for $C=1$ (black) and $C=1.8$
(green).}
\label{fig:nonfwdP}
\end{center}
\end{figure}

\begin{figure}[h!]
\begin{center}
\includegraphics[width=0.66\textwidth]{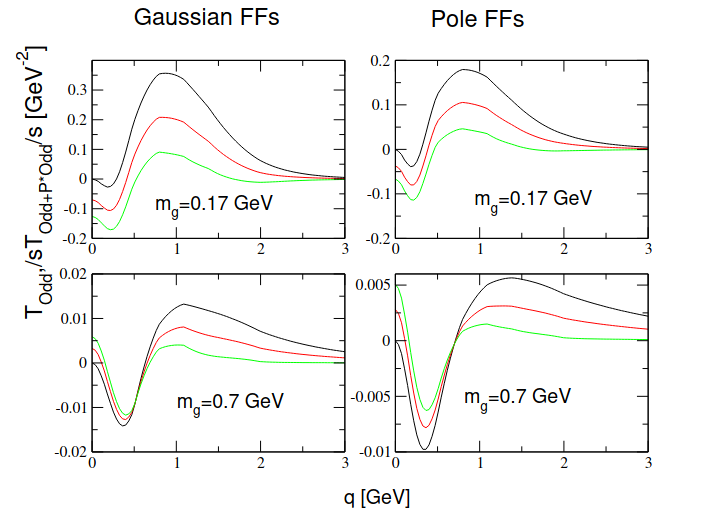}
\caption{Odderon amplitudes for the non-forward regeneration of \KS{} on
protons, using two values of the effective gluon mass $m_g$ and the
{\it Gaussian} and {\it pole} form factors: the black curves give
$T_{\rm Odd}/s$, while the red and green curves show
$(T_{\rm Odd}+T_{P\otimes{\rm Odd}})/s$ for $C=1$ (red) and $C=1.8$
(green).}
\label{fig:Zakh3}
\end{center}
\end{figure}

\end{document}